\documentstyle[12pt]{article}

\def\.{\!\cdot\!}
\def\:{\cdots}
\def\[{\left[}
\def\]{\right]}
\def\({\left(}
\def\){\right)}

\def\g2{{g^2\over 2\pi}}

\def\r2{\sqrt{2}}

\begin{document}
\input epsf.tex

\rightline{McGill/96-23}
%96-06-16
\vspace{.5 cm}
\begin{center}
{\Large\bf Nonabelian Cut Diagrams and Their Applications$^*$\\}
\vspace{.5 cm}
{C.S. Lam$^{\dag}$}\\
\bigskip
{\it Department of Physics, McGill University,\\
3600 University St., Montreal, P.Q., Canada H3A 2T8}
\end{center}

\begin{abstract}
A new kind of cut diagram is introduced to sum Feynman diagrams with
nonabelian vertices. Unlike the Cutkosky diagrams which compute the
discontinuity of single Feynman diagrams, the nonabelian cut diagrams
represent a resummation of both the real and the imaginary parts of
Feynman diagrams related by permutations. Several applications of the
technique are reported, including a resolution of the apparent inconsistency
of the baryon problem in large-$N_c$ QCD, a simplified calculation of
high-energy low-order QCD diagrams, and progress made with this technique
on the unitarization of the BFKL equation.
\end{abstract}

\section{Nonabelian cut diagrams}
This note serves to introduce the recently developed nonabelian cut diagrams
\cite{1,2} and to summarize their applications to date \cite{2,3,4}.

A tree diagram of the type shown in Fig.~1, with bosons of relatively small
momenta $q_i$ absorbed or emitted  from a  particle of large momentum $p$,
has an amplitude given approximately by
\begin{eqnarray}
-2\pi i\delta\(\sum_{j=1}^n\omega_j\)
\(\prod_{i=1}^{n-1}{1\over \sum_{j=1}^i\omega_j+i\epsilon}\)\.t_1t_2\:t_n\.V
\equiv a[12\:n]\.t[12\:n]\.V\ ,
\end{eqnarray}
where $\omega_j=2p\.q_j$, $t[12\:n]\equiv t_1t_2\:t_n$, with
$t_i$ being some nonabelian vertices. $V$ contains other objects
such as coupling constants, etc., independent of the ordering of the bosons.
For easy reference we shall assume $p$ to be the momentum of a fermion,
though the kinematical and combinatorial formulas discussed below remain
unaltered whatever that particle is.

\begin{figure}
\vskip -0 cm
\centerline{\epsfxsize 3 truein \epsfbox {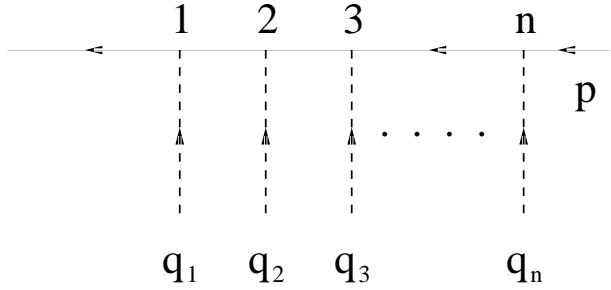}}
\nobreak
\vskip 0 cm\nobreak
\vskip .1 cm
\caption{A tree diagram with $n$ bosons emitted or absorbed.}
\end{figure}

The momenta $q_i$ have been assumed in (1.1) to be much smaller than the
momentum $p$, so that the approximation $(p+\sum_jq_j)^2-M^2\simeq 2p\.
\sum_jq_j$ can be used to compute the denominators of propagators. 
This would be so in the presence of a very large fermion mass,
as is the case for baryons in large-$N_c$ QCD. It is also valid when the
fermion is taken from the incoming beam of a high energy process like
quark-quark scattering. These two applications will be discussed separately
in the next two sections.

Our aim is to compute the sum of the nonabelian amplitude
(1.1) over the $n!$ permuted tree
diagrams, with the help of two exact combinatorial formulas involving
$a[\:]$ and $t[\:]$. Since the momenta $q_i$ are allowed to be offshell,
these formulas are equally valid when the tree in Fig.~1 is part of a
much larger loop diagram, so the formalism can be used just as well to compute
sums of loop diagrams with the boson lines so permuted.

To state these formulas we must first introduce suitable notations
to describe and to manipulate these tree diagrams.
We will use $[\sigma]=[\sigma_1\sigma_2\:\sigma_n]$ to denote
a tree diagram whose bosons are numbered in the order
$\sigma_1, \sigma_2, \:, \sigma_n$, from left to right. 
Hence Fig.~1 is $[12\:n]$.
If $[T_i]$ are tree diagrams, then $[T_1T_2\:T_A]$ is taken to mean
the tree diagram obtained by merging these $A$ trees together in that order. 
For example, if $[T_1]=[123]$
and $[T_2]=[45]$, then $[T_1T_2]=[12345]$. The notation $\{T_1;T_2;\:;T_A\}$,
on the other hand,
is used to denote the {\it set} of all tree diagrams obtained by
{\it interleaving} the trees $T_1, T_2, \:, T_A$ in all possible ways.
This set contains
$(\sum_a n_a)!/\prod_an_a!$ trees if $n_a$ is the number of boson lines of
tree $T_a$. In the example above, $\{T_1;T_2\}$
contains the $5!/3!2!=10$ trees [12345], [12435], [12453], [14235],
[14253],
[14523], [41235], [41253], [41523], and [45123].

Correspondingly,
$a\{T_1;T_2;\:;
T_A\}$ is defined to be the sum of the amplitudes $a[T]$ for every tree $T$
in this set.

With these notations, we can proceed to state the two combinatorial
theorems. The {\it factorization formula} \cite{1} asserts that the sum of
amplitudes on the left of the following equation 
factorizes into the expression on the right:
\begin{eqnarray}
a\{T_1;T_2;\:;T_A\}=\prod_{a=1}^Aa[T_a]\ .
\end{eqnarray}
In the special case when each $[T_i]=[i]$ is just a vertex, the set
$\{1;2;\:;A\}$ consists of the $A!$ permuted trees
of $[12\:n]$,
and the factorization formula is simply
the well-known eikonal formula \cite{5}
\begin{eqnarray}
a\{1;2;\:;A\}=\prod_{a=1}^A\[-2\pi i\delta(\omega_a)\]\ .
\end{eqnarray}

The factorization formula is used to show reggeized factorization \cite{4}.
It is also applied to derive the {\it multiple commutator formula} \cite{1}
used to sum up the $n!$ permuted trees of the nonabelian amplitude (1.1):
\begin{eqnarray}
\sum_{\sigma\in S_n}a[\sigma]t[\sigma]=\sum_{\sigma\in S_n}a[\sigma]_c
t[\sigma]'_c\ .
\end{eqnarray}
The symbols and meaning of this equation will be explained below.

To each Feynman tree diagram $[\sigma]=[\sigma_1\sigma_2\:\sigma_n]$
we associate
a {\it cut diagram} $[\sigma]_c$ by putting cuts on specific fermion
propagators as follows. Proceeding from left to
right, a cut is put after a number if and only if a smaller number does not
occur to its right. An external line is considered to be equivalent
to a cut so there is never an explicit cut at the end of the tree.

Designating a cut by a vertical bar behind a number,
here are some examples of where cuts are to be put into Feynman trees:
$[1234]_c=[1|2|3|4]$, $[3241]_c=[3241]$, and $[2134]_c=[21|3|4]$.

The {\it complementary cut
diagram} $[\sigma]'_c$ is one where lines cut in $[\sigma]_c$ are not cut in
$[\sigma]'_c$, and vice versa. Thus $[1234]'_c=[1234]$,
$[3241]'_c=[3|2|4|1]$, and $[2134]'_c=[2|134]$.

The cut amplitude $a[\sigma]_c$
is the Feynman amplitude $a[\sigma]$
with the propagator at a cut taken to be the Cutkosky cut propagator
$-2\pi i\delta(\sum_j\omega_j)$
instead of the Feynman propagator $(\sum_j\omega_j+i\epsilon)^{-1}$.
However, cuts are placed here only on
high speed fermion lines, and as (1.4) indicates, these {\it 
nonabelian cut diagrams}
represent a resummation and not a discontinuity as is the case in Cutkosky
cut diagrams.

The nonabelian quantum number
$t[\sigma]'_c$ is determined from the complementary cut diagram $[\sigma]'_c$
by replacing the product of
nonabelian vertices separated by cuts with their commutators. For example,
$t[1234]'_c=t[1234]=t_1t_2t_3t_4$,
$t[3214]'_c=t[3|2|4|1]=[t_3,[t_2,[t_4,t_1]]]$,
and $t[2134]'_c=t[2|134]=[t_2,t_1]t_3t_4$.

This completes the description and the explanation of eq.~(1.4). Eq.~(1.4)
reduces to (1.3) when all the $t_a$ commute, so it
can be considered as a nonabelian generalization of the eikonal formula.

Physically we may think of 
the eikonal formula
as exhibiting a very interesting interference phenomenon. According to (1.3),
as a result of the interference of the $A!$ amplitudes, a very strong
$A$-dimensional peak is found at all $\omega_a=0$. At any other energy
the amplitude vanishes by destructive interference. The nonabelian version
(1.4) is more subtle because it does different things to different nonabelian
channels. The peak at $\omega_a=0$ is generally weaker, being only of $B\le A$ 
dimensions, but this is compensated by having $A-B$ commutators
of the nonabelian vertices. The exact physical significance of the
commutators depends
somewhat on the application and will be discussed separately in the next
two sections.

\section{The baryon problem in large-$N_c$}

Suppose Fig.~1 represents meson-baryon inelastic
scattering in large-$N_c$ QCD.
The Yukawa coupling constant $g$ at each vertex is known to be proportional to
$\sqrt{N_c}$, hence the tree diagram will grow like $N_c^{n/2}$. On the other
hand, the complete tree amplitude after summing over the 
the $n!$ permuted diagrams is known to behave like $N_c^{1-n/2}$, so there is
a discrepancy of $n-1$ powers of $N_c$ between individual Feynman diagrams
and their sum. For the meson-baryon inelastic scattering problem at large
$N_c$ to be consistent, a strong cancellation must occur between the individual
diagrams, and more so for larger $n$.

The multiple commutator formula (1.4) can be used to demonstrate that this
indeed happens \cite{3,6}, 
so the large-$N_c$ baryon problem is indeed self consistent.
Essentially, what happens is that each time a commutator appears, 
the $N_c$ power is reduced by 1. So in the term where the {\it complementary
cut diagram} has a cut in every baryon propagator, a reduction
by a factor of $N_c^n$ occurs, which is 
precisely what is required to achieve the cancellation
needed for the consistency.  Note that in this case the 
corresponding {\it cut diagram}
contains only Feynman propagators. 
All the other terms in (1.4) have at least one cut propagator in
the {\it cut diagram}, so the amplitude $a[\sigma]_c$ would contain
at least one $\delta(\omega_a)$ which is zero in the generic energy
configuration where all the meson energies are non-vanishing. By analytic
continutation, we can now assert that the cancellation always occurs and the
baryon problem is self-consistent at all energies.

\section{High-energy QCD scattering}
Consider quark-quark elastic scattering in an $SU(N_c)$
theory where the `quarks' are allowed to carry any color. In this problem
$N_c$ could be 3 and is not necessarily large. We are interested in
the situation where the energy $\sqrt{s}$ is much larger than the momentum
transfer $\sqrt{-t}$. Since high-energy cross-sections are spin independent,
what is being discussed below applies to gluon-gluon scattering as well.

Unlike the situation of last section where terms containing a 
$\delta(\omega_a)$ may be discarded, the destructive interference
exhibited here is much more subtle. In a loop amplitude, the
contribution from the $\delta$-function must be retained, but it turns out that
the amplitude would be a factor of $\ln s$ down in
the presence of each $\delta(\omega_a)$ \cite{2,4}.

The role of the nonabelian quantum numbers $t_a$, in this case color,  
is also different. The commutators now specify in which color channels
these interference effects are to be seen. For example, the cut diagram
without any cut has no interference suppression in its amplitude.
The quantum number of that term is given by a multiple commutator of
$t_a$, corresponding to the adjoint color channel, so in $SU(3_c)$ the
octet channel remains intact.
The amplitude in any other color channel has at least one $\delta(\omega_a)$
and will be suppressed.

It can be shown \cite{4} that such automatic suppression in the
nonabelian cut diagrams does not occur in 
Feynman diagrams. Individual
Feynman diagrams would bear a larger power of $\ln s$ in these channels, 
and the suppression would occur only when the individual
Feynman diagrams are added together \cite{2,4,7}. This means that
contributions to such color channels can be obtained only when individual
Feynman diagrams are computed to subleading orders---a very
difficult task in general. In contrast, this is not be a problem
in nonabelian cut diagrams so they may be computed just to leading-log
accuracies. This is a great advantage especially for high-order calculations.

The gluon in QCD scattering is known to be reggeized, and the Low-Nussinov
Pomeron \cite{8} appears in the two-reggeon-exchange amplitude.
A leading-log computation of this Pomeron \cite{9} leads to a total
cross-section growing like a positive power of $s$,  violating 
the Froissart bound.
Subleading logarithms are therefore needed to restore unitarity
to the BFKL equation \cite{9}, and these
may come from multi-reggeon exchanges [10]. What is lacking is the proof that
the sum of Feynman diagrams indeed factorizes into multi-reggeon amplitudes.
To be sure this factorization has been verified in low-order calculatiuons, 
completely to the 6th order and partially
to the 8th and 10th orders \cite{10}.
An attempt to extend this to all orders in the usual approach would be
extremely difficult, on account of the delicate cancellations 
discussed in the last paragraph, and because it is difficult to see
why the sum of the complicated diagrams with all the criss-crossing of lines
should {\it factorize}. Nonableian cut diagrams are potentially capable
of solving both of these difficulties, for delicate cancellations plaguing
sums of Feynman diagrams do not occur here, and the factorization formula
(1.2) which led to their derivation can also be used to obtain reggeon 
factorization. Accordingly we have initiated a program to use the nonabelian
cut diagrams to study the validity of this multi-reggeon factorization.
We are now able to prove it to be true for
$s$-channel-ladder diagrams to all orders \cite{4}. More work is required
for more complicated diagrams.

\section{Acknowledgements}
I wish to thank my collaborators Y.J. Feng, O. Hamidi-Ravari, and K.F. Liu.
This research is supported in part by the by the Natural Science and
Engineering Research Council of Canada, and the Fonds pour la
Formation de Chercheurs et l'Aide \`a la Recherche of Qu\'ebec.

\end{document}